\documentclass[twocolumn]{emulateapj}

\newcommand{\Lsun}{$L_{\odot}$}
\newcommand{\Msun}{$M_{\odot}$}

\newcommand{\Mdot}{$\dot{M}$}

\shorttitle{Dynamical processes with dust mineralogy}
\shortauthors{McClure et al.}

\begin{document}

\title{Probing dynamical processes in the planet forming region with dust mineralogy}


\author{M. K. McClure\altaffilmark{1,2}, P. Manoj\altaffilmark{3}, N. Calvet\altaffilmark{1}, L. Adame\altaffilmark{1}, C. Espaillat\altaffilmark{4}, D. M. Watson\altaffilmark{3}, B. Sargent\altaffilmark{6}, W. J. Forrest\altaffilmark{3}, P. D'Alessio\altaffilmark{7}}

\altaffiltext{1}{Department of Astronomy, The University of Michigan, 500 Church St., 830 Dennison Bldg., Ann Arbor, MI 48109; melisma@umich.edu, ncalvet@umich.edu, adamel@umich.edu}
\altaffiltext{2}{NSF Graduate Research Fellow}
\altaffiltext{3}{Department of Physics and Astronomy, University of Rochester, Rochester, NY 14627, USA; manoj@pas.rochester.edu, dmw@pas.rochester.edu, forrest@pas.rochester.edu}
\altaffiltext{4}{Center for Astrophysics, 60 Garden Street, Cambridge, MA 02138, USA; cespaillat@cfa.harvard.edu}
\altaffiltext{5}{NSF Astronomy and Astrophysics Postdoctoral Fellow}
\altaffiltext{6}{Center for Imaging Science and Laboratory for Multiwavelength Astrophysics, Rochester Institute of Technology, 54 Lomb Memorial Drive, Rochester, NY 14623, USA; baspci@rit.edu}
\altaffiltext{7}{Centro de Radioastronom\'{i}a y Astrof\'{i}sica, Universidad Nacional Aut\'{o}noma de M\'{e}xico, 58089 Morelia, Michoac\'{a}n, M\'{e}xico; p.dalessio@crya.unam.mx}

\begin{abstract}
We present {\it Herschel} Space Observatory \footnote{Herschel is an ESA space observatory with science instruments provided by European-led Principal Investigator consortia and with important participation from NASA.}  PACS spectra of GQ Lup, a protoplanetary disk in the Lupus star-forming region.  Through SED fitting from 0.3$\mu$m to 1.3 mm, we construct a self-consistent model of this system's temperature and density structures, finding that although it is 3 Myr old, its dust has not settled to the midplane substantially.  The disk has a radial gradient in both the silicate dust composition and grain size, with large amorphous grains in the upper layers of the inner disk and an enhancement of submicron, crystalline grains in the outer disk.  We detect an excess of emission in the {\it Herschel} PACS B2A band near 63$\mu$m and model it with a combination of $\sim$15 to 70$\mu$m crystalline water ice grains with a size distribution consistent with ice recondensation-enhanced grain growth and a mass fraction half of that of our solar system.  The combination of crystalline water ice and silicates in the outer disk is suggestive of disk-wide heating events or planetesimal collisions.  If confirmed, this would be the first detection of water ice by {\it Herschel}.
\end{abstract}


\section{Introduction}

While they are a small fraction by mass of material in the interstellar medium (ISM) and the primordial disks around pre-main sequence stars, dust grains play pivotal roles in the physical processes that shape these disks.  The optical properties of the dust dictate the structure of the disk \citep[e.g.][]{dalessio+06}, and spectral features can be used to probe dynamical processes such as settling, turbulence, and grain growth in the disk atmosphere \citep{bouwman+08,watson+09,turner+10}.  

In addition to the current disk structure, the grains' composition, structure, and size distribution record the local thermal history and kinetic conditions of the regions in which they formed.  For example, silicate dust grains are at least 98\% amorphous in the ISM \citep{kemper+04} but are found to be partly crystalline in protoplanetary disks \citep{bouwman+03,sargent+09b}.  Crystallization is highly sensitive to the specific local thermal conditions; different silicate stoichiometries form at different temperatures and pressures by thermal annealing or sublimation and re-condensation \citep[][and references therein]{hm09}.  Both processes occur at temperatures greater than $\sim$ 700K.  Nominally, this temperature is characteristic of the inner 1AU of a typical disk surrounding a T Tauri star (TTS).  However, crystalline silicates are observed out to 40AU in solar system comets \citep{crovisier+97} and some TTS systems \citep{espaillat+07}. Either transportation from the inner 1AU to the outer disk via winds, turbulent diffusion, or radial flows \citep{gail01, ciesla09b}  or {\it in situ} formation by heating events above the 700K threshold (shocks or planet destruction) \citep{harkerdesch02} are required to explain these results. Characterizing the mineralogical zoning of TTS disks at different radii is necessary to discriminate between these different scenarios.

Another major component of dust in disks is water ice, which may play an important role in forming planetesimals, either by helping grains stick together \citep{ormel+11, ks11} or by increasing the dust to gas mass ratio at the snowline, which
may induce the accumulation of grains at this location \citep{kl07}.  Although ice has been detected in absorption at NIR wavelengths in highly inclined disk systems \citep{pontoppidan+05}, and hot water vapor has been detected in MIR {\it Spitzer} spectra \citep[e.g.][]{cn08}, these observations probe mainly the upper layers of the disk, leaving the distribution and phase of water in the outer disk and midplane largely unknown.  Recent detections of cold water vapor in TW Hya require much less water than indicated by MIR, suggesting that icy grains may settle to the midplane leaving dry grains in the disk atmosphere \citep{hogerheijde+11}.  Identifying where and how much ice is present in TTS systems would help resolve this question.

With this motivation, we investigate GQ Lup, a 3 Myr old \citep{sd+08} protoplanetary K7 TTS system in the Lupus molecular cloud complex with a sub-stellar mass companion at 0\farcs7 separation \citep{neuhauser+05} and a circumprimary disk that appears to be truncated \citep{dai+10}. Given the range of inclination angles found for the system, $\sim$30$^{\circ}$ to 50$^{\circ}$ \citep[hereafter SD08]{broeg+07,sd+08}, the separation amounts to 120 to 160AU at the distance of Lupus, 150 pc \citep{franco02}.   Using {\it Herschel} PACS, we obtained 55 to 145$\mu$m spectra of GQ Lup.  To characterize simultaneously the distribution of silicates and water ice in relation to the disk structure, we combined these data with archival {\it Spitzer} spectroscopy and ancillary photometry and used irradiated accretion disk models to fit the spectral energy distribution (SED) of GQ Lup.

\section{Observations and data reduction}
\label{obsred}

We observed GQ Lup using {\it Herschel} \citep{pilbratt+10} on 8 January, 2012 (OBSID 1342238375) with PACS \citep{poglitsch+10} range spectroscopy modes B2A (51-73$\mu$m) and R1S (102-145$\mu$m) at Nyquist-sampling (R$\sim$1500) and a total time of 7774 seconds.  The data were reduced using the standard data reduction pipeline in HIPE version 9.0 \citep{ott2010}.  We extracted the spectra from each spaxel, confirmed that the source was point-like and well centered on the central spaxel within the pointing uncertainty of $\sim$2$^{''}$, and applied the PSF correction to the central spaxel spectrum.  The uncertainty in PACS absolute flux calibration can be up to 30\%; however, GQ Lup was observed by both IRAS, at 60 and 100$\mu$m, and AKARI, at 65 and 90$\mu$m.   We use this photometry to confirm the absolute photometric accuracy of the PACS spectrum.  The point-to-point variation of the spectrum after rebinning by a factor of 10 is $\sim$15\%; we assume this as our relative spectral uncertainty.

The {\it Spitzer} IRS \citep{houck+04} low (SL, 5$-$14$\mu$m, $\lambda/\Delta\lambda$=60$-$120, AORID 5644032) and high (SH, 10$-$19$\mu$m, LH, 19$-$35$\mu$m, $\lambda/\Delta\lambda$=600, AORID 27064576) spectral resolution data were observed on 30 August 2004 and 2 September 2008 as part of programs 172 and 50641, respectively.  We reduced them with SMART \citep{higdon+04} in the same way as in \citet{mcclure+10}, with the exception that the SH/LH data were sky subtracted from off-source frames included in that AOR.  We estimate the spectrophotometric uncertainty to be $\sim$5\%.

\section{Analysis}
\label{allanalysis}

\begin{figure}
\includegraphics[angle=0, scale=0.9]{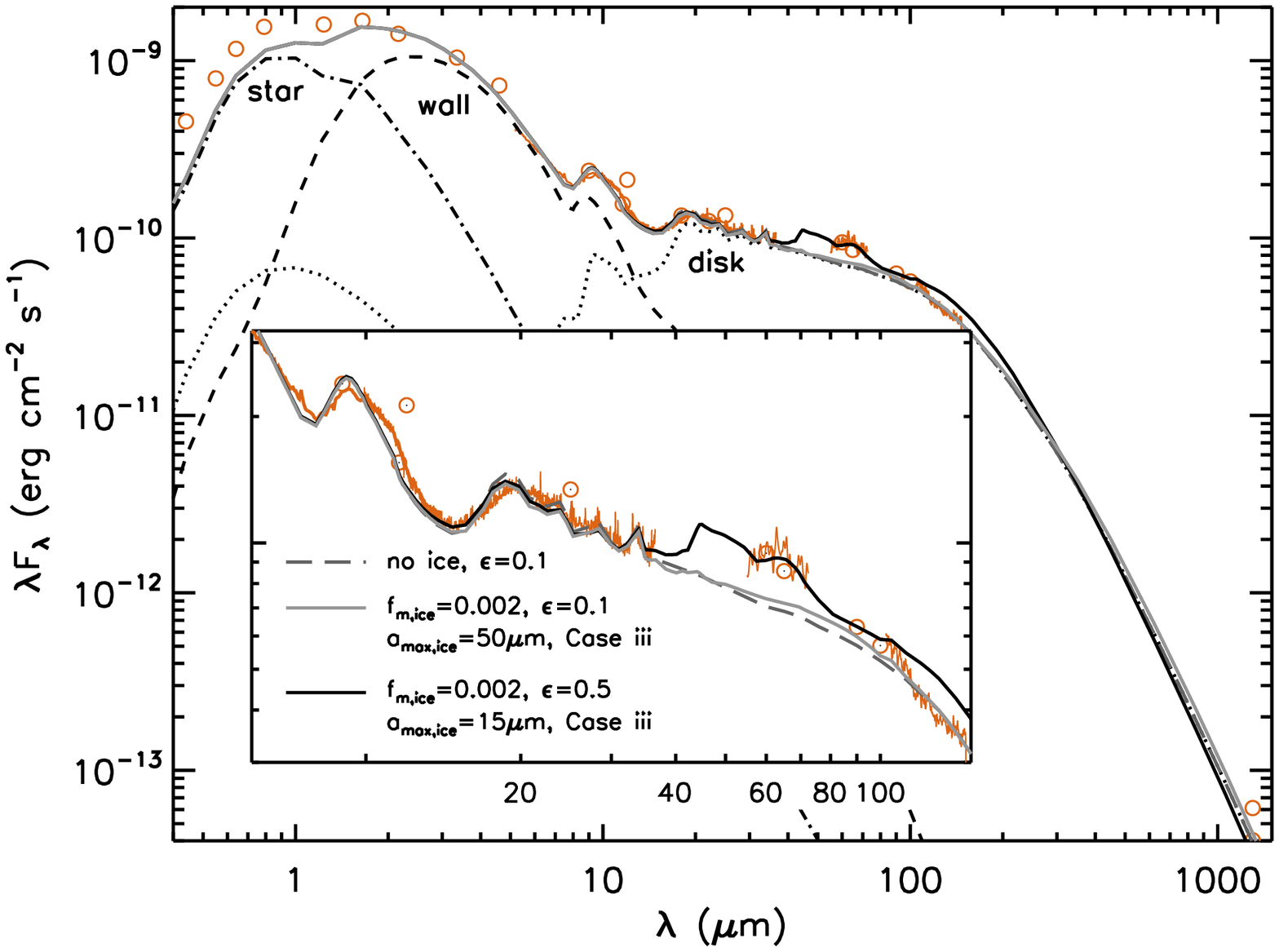}
\caption{SED (orange lines and symbols) for GQ Lup.  Photometry are from \citet{covino+92}, 2MASS, WISE, AKARI, IRAS, and \citet{dai+10}.  Spectra are from the {\it Spitzer} Heritage Archive and this work.  The best fitting non-ice model is shown, along with two ice models.  One fits everything but B2A (50$\mu$m grains, solid grey) and the other fits everything except 120-140$\mu$m (15$\mu$m grains, solid black).  The remaining model parameters are given in Table \ref{sampletab}.  The model does not fit the optical data because we do not include emission from the accretion shock itself.       \label{gqbestfit}}
\end{figure}

\begin{figure}
\includegraphics[angle=0, scale=0.9]{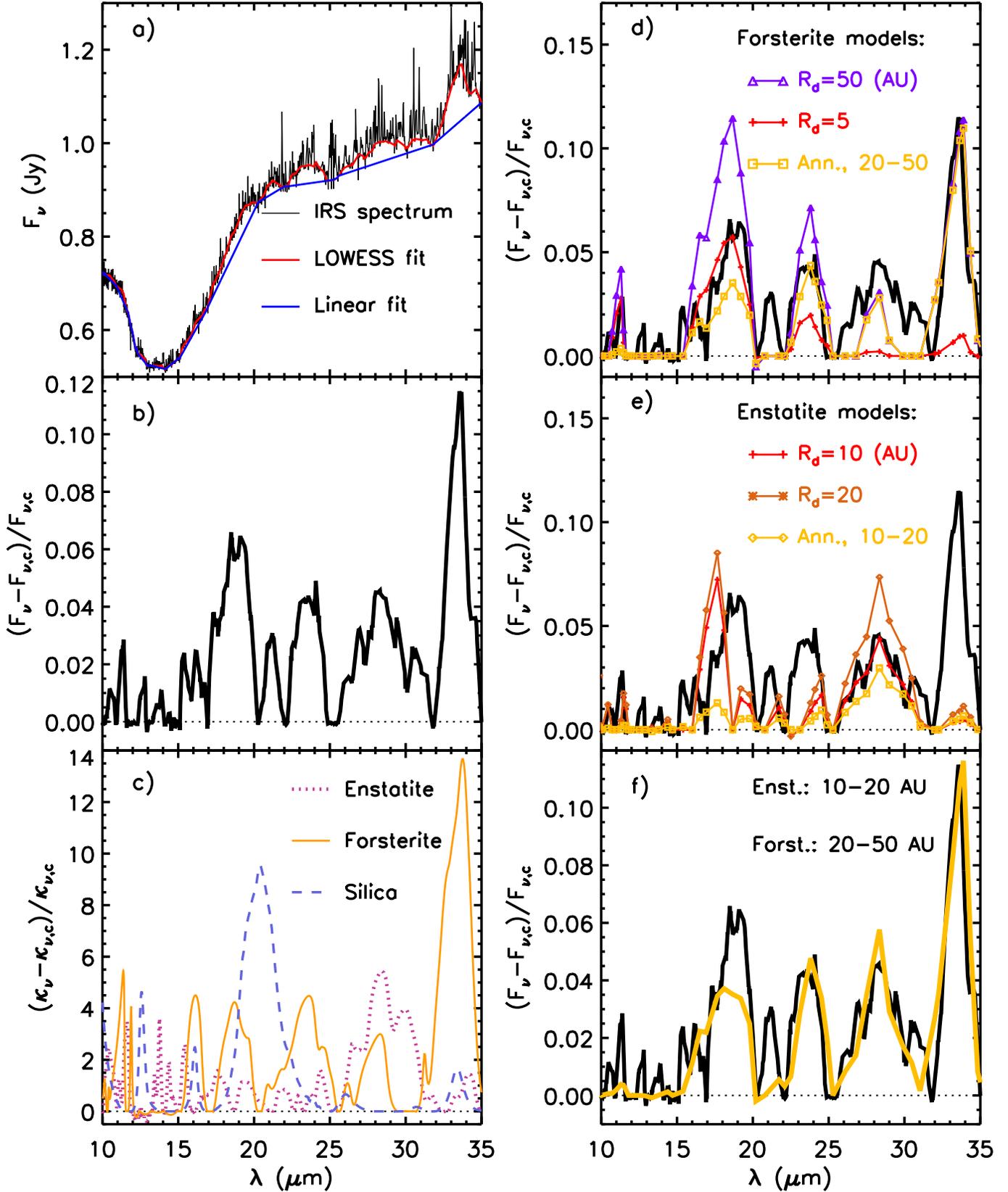}
\caption{{\it a)} Method for isolating crystalline silicate features in the IRS spectrum. {\it b)} Observed crystalline features. {\it c)} Opacities for three most common crystalline silicates.  {\it d)} Model forsterite features for fixed forsterite mass fraction and varied outer disk radius, as well as an annulus.  {\it e)} Model enstatite features with best-fitting annulus.  {\it f)}  Best-fitting combination of annuli, 18\% forsterite and 15\% enstatite.  The poor match to the 18$\mu$m region is due to fitting a linear baseline to a region that has intrinsic curvature. \label{forstrdfit}}
\end{figure}

The SED of GQ Lup, is shown in Figure \ref{gqbestfit}.  It has a strong excess at all infrared wavelengths, indicating the presence of a dust sublimation wall and disk.  However, the disk emission drops off rapidly with increasing wavelength, consistent with the conclusion by \citet{dai+10} that it is outwardly truncated.  The {\it Herschel} B2A spectrum shows a peaked triangular shape around 63$\mu$m suggestive of the water ice feature located there.  We see no evidence for a forsterite feature at 69$\mu$m.  In the IRS spectrum, we identify the major crystalline features by fitting a non-parametric locally weighted scatterplot smoothing (LOWESS) baseline to the data, taking this as the `dust continuum' beneath the molecular lines, and subtracting a linear fit to regions between known crystalline silicate features to the IRS spectrum (Fig. \ref{forstrdfit}a). There are strong forsterite features at 23 and 33$\mu$m, blended forsterite-enstatite features at 18 and 28$\mu$m, and weak enstatite features around 11$\mu$m.

To determine the composition and structure of the disk, we construct temperature and density structures using the \citet{dalessio+06} irradiated accretion disk models, which assume the disk is heated by stellar irradiation and viscous dissipation.  Steady accretion and viscosity are parametrized through constant \Mdot and $\alpha$, respectively \citep{shakurasunyaev73}. The disk consists of gas and dust, the latter of which is comprised of two grain populations mixed vertically.  Settling is parameterized through $\epsilon=\xi/\xi_{standard}$, where the denominator is the sum of the mass fraction of the different components relative to gas and the numerator is the mass fraction in the small dust population.  

The silicate and graphite grains have size distributions $n(a)=n_0a^{-3.5}$, where $a$ is the grain radius with limits of 0.005$\mu$m and $a_{max}$.  To test whether the ice grains have grown larger than the silicate grains, we consider three size distributions:  (Case $i$) the same power law dependence and $a_{max}$ as the silicate and graphite grains; (Case $ii$) the same power law dependence but a larger $a_{max}$; (Case $iii$) a power law dependence of $n(a)=n_0a^{2.0}$ \citep[e.g.][]{ks11}.  In this model, ice exists everywhere below the thermal sublimation temperature; while other processes, e.g. photodesorption, are likely to modify this snowline \citep{oberg+09c}, a self-consistent treatment is beyond the scope of this letter and reserved for future work.

We compute opacities for the graphite and water ice grains using optical constants from \citet{dl84} and \citet{warren84} and Mie theory, assuming that the grains are segregated spheres \citep{pollack+94}.  Silicates are divided into amorphous and crystalline versions of two stoichiometries: olivines ($Mg_{2-2x}Fe_{2x}SiO_4$) and pyroxenes ($Mg_{1-x}Fe_{x}SiO_3$), where $x=Fe/(Fe+Mg)$ indicates the iron content.  Opacities for the amorphous olivine and pyroxene are computed with optical constants from \citet{dorschner+95} that have $x=0.5$ for olivine and range from $0.05$ to $0.6$ for pyroxene.  The opacities for crystalline olivine (forsterite) and pyroxene (enstatite) are taken directly from those calculated by \citet{sargent+09b}.
 
We implement a vertical dust sublimation wall with an atmosphere following the prescription of \citet{dalessio+04}.  Dust is present once the disk temperature drops below the dust sublimation temperature, and this radius, $R_{wall}$, defines the radial location of the beginning of the wall atmosphere, where the dust is optically thin.  We allow the dust properties of the wall to vary independently of those in the disk to simulate the effects of a radial gradient in the inner disk mineralogy; in particular, we vary the iron content of the silicates in the wall, while in the disk we assume the iron content of the best-fitting amorphous olivine and pyroxene determined by \citet{sargent+09b} from comparison with disks in Taurus.

As input to the code, we assume the stellar and accretion properties within the range given by SD08 (values listed in Table \ref{sampletab}).  They measured high veiling in the system over the course of a month and estimated mass accretion rates between 10$^{-8}$ and 10$^{-7}$ \Msun/yr from comparing their observed emission lines with magnetospheric accretion models; we adopt an intermediate value of 5$\times$10$^{-8}$ \Msun/yr. The outer radius is not totally unconstrained, as the mass ratio range of the companion is 0.014 to 0.03, which is too large for the circumprimary disk to radially extend over the 1:2 binary orbital resonance \citep{al94}.  Therefore the disk can be at most 70AU in radius.  

We create a grid of $\sim$350 models varying $\epsilon$, $\alpha$, $a_{max,s}$, the ice mass fraction ($f_{m,i}$), silicate composition, and $R_d$ and minimize the reduced $\chi^2$ statistic, $\chi_r^2$, between the models and observations to determine the best fit.  While there are degeneracies between $\epsilon$, $\alpha$, $f_{m, i}$, and $R_d$, the large wavelength coverage of our SED allows us to break some of them.  For example, only $\epsilon\ge0.1$ can provide enough continuum emission to fit the IRS spectrum, while $\epsilon\le0.5$ fits the slope between PACS R1S and the 1.3 mm photometry.  Likewise, smaller disks (e.g. $R_d\sim$20AU) underproduce the 20 to 35$\mu$m absolute flux while larger disks (e.g. $R_d\sim$70AU) overproduce the 100 to 140$\mu$m flux.  Once these parameters have been constrained, this implies an $\alpha$ between 0.1 and 0.01.  However, since the fit to $\alpha$ comes mainly from the submillimeter flux, which depends on the surface density, $\Sigma\propto$\Mdot/$\alpha$, the estimated range of \Mdot yields 0.02$\leq\alpha\leq$0.2 for the same fit to the SED.

\begin{figure}
\includegraphics[angle=0, scale=0.9]{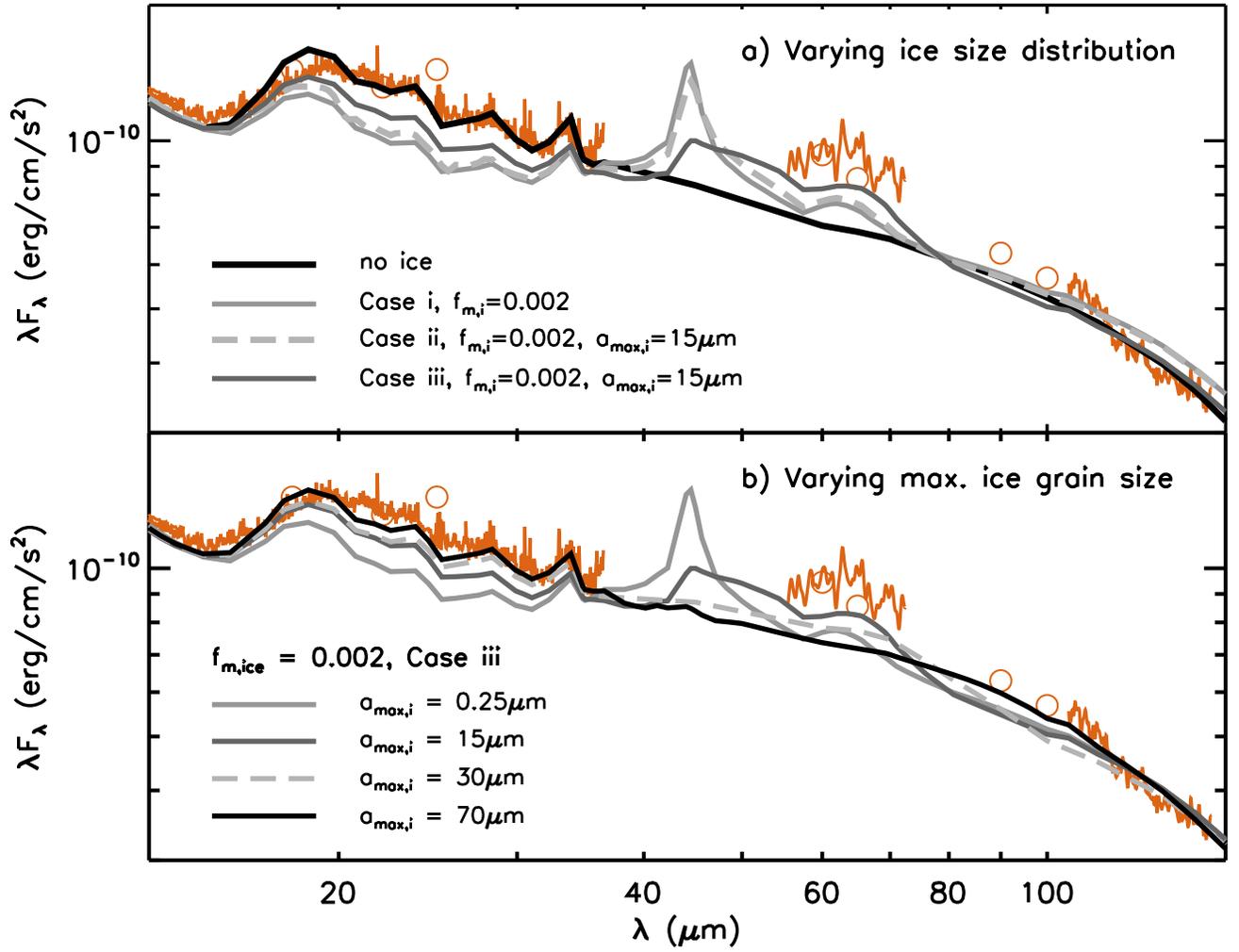}
\caption{Effects of varying ice grain properties, i.e. mass fraction and size distributions (panel a) and maximum grain size (panel b), for a fixed disk structure and silicate/graphite dust properties. \label{varyice}}
\end{figure}

Models without ice fail to reproduce either the shape or absolute flux of the PACS B2A spectrum; in Fig \ref{varyice}a we show this model and how the SED changes when ice with the distributions given by the previously defined size distributions Case $i$, $ii$, and $iii$ is added.  Case $i$ lowers the flux too much over the {\it Spitzer} range and fails to produce enough flux over B2A.  Case $ii$ produces only a slight increase in the flux between 55 and 70$\mu$m.  With Case $iii$, we are able to produce enough flux over both the IRS and PACS range.  For Case $iii$, the B2A and R1S spectra are better reproduced by different $a_{max,i}$, 15$\mu$m and 50 to 70$\mu$m, respectively (Fig. \ref{varyice}b).  The mass fraction of ice that produces the best fit to the data for this disk structure is 2$\times$10$^{-3}$ with respect to gas, or $\sim$25\% of the total dust content, which is half of the solar composition \citep{pollack+94}.  Although our model includes ice at this abundance in the midplane as well, because the disk has so little dust settling, we would require spatially resolved submillimeter photometry to test whether the midplane ice abundance differs from that of the upper layers.

The best fitting wall has an $a_{max}$ of 3$\mu$m and is comprised of 100\% glassy pyroxene with Fe/(Fe+Mg)=0.4.   With the assumption of a radially uniform disk composition, the best-fitting models require silicate grains with $a_{max}$=0.25$\mu$m and a mixture of 80\% amorphous silicates, of which 5\% are olivine and 75\% pyroxene (Fe/(Fe+Mg)=0.2), and 20\% crystalline silicates, evenly split between forsterite and enstatite.   Although our model does not yet include radial composition gradients, we can still use it to probe the location of the crystalline silicates.  We compute a grid of theoretical disks truncated outwardly at 5AU intervals from 50AU to 5AU for compositions of pyroxene-forsterite and pyroxene-enstatite, allowing the individual crystalline silicate fractions to vary between 5\% and 30\%.  By isolating the model forsterite features in the same manner as for the IRS spectrum, we see that the models with sufficient forsterite abundance to fit the  33$\mu$m feature produce too much flux in the 23$\mu$m features (Fig. \ref{forstrdfit}) and {\it vice versa}, indicating that forsterite is not uniformly distributed throughout the disk. 

To compare the data with the emission from an annular region of the disk, we subtract the emission of the other radii from the 50AU disk and compare the ratio of the 23 to 33$\mu$m feature to determine which emitting annulus is the best match to our data.  Forsterite distributed between 20 and 50AU fits the relative fluxes best and implies a forsterite silicate fraction of $\sim$18\% in this region. A match to the observed enstatite emission of an annulus of 10 to 20AU with a fraction of 15\% is obtained in the same way.  We note that the non-detection of the 69$\mu$m forsterite feature in the PACS spectrum places an upper limit of 90\% on the forsterite fraction in the outer disk upper layers. To cross-check our results, we perform a two-temperature spectral decomposition \citep[e.g.][]{sargent+09b} on the IRS spectrum.  Its results confirm that the inner disk is dominated by large, amorphous pyroxene grains with no crystalline grains detected at $>1\sigma$.  The outer disk, however, contains small pyroxene grains with forsterite and enstatite present at $3\sigma$ and $2\sigma$, respectively.  The resulting system properties, assuming radially constant dust abundances in the disk, are listed in Table \ref{sampletab}.   The final best fit has $\chi_r^2$=1.2, and is displayed in Fig. \ref{gqbestfit}.

\section{Discussion and conclusions}
\label{discussion}
\subsection{An iron-rich and turbulent inner disk?}
The radius of the wall is proportional to the ratio of a grain's opacity weighted by the stellar and local temperature and the assumed dust sublimation temperature.  Of the grains we tested, grains with $x$=0.4 (40\% Fe content) had the largest opacity ratio, so they absorb heat more efficiently than they emit it and achieve the dust sublimation temperature at larger radii than grains with less iron.  For a wall whose height is given in a fixed number of gas pressure scale heights, the solid angle to the observer is then larger, and therefore the peak observed flux is greater.  The high iron content of our inner disk is consistent with the Fe/Mg $\sim$1 ratio found, on average, for pre-solar silicate grains in meteorites \citep{nguyen+10}.  However, this work does not explore the possibility that other near-infrared continuum sources that might be expected in the inner disk, e.g. metallic iron which represents 60\% of Mercury's total mass, could be present in the wall.  The large maximum grain size of the wall, taken in conjunction with the high \Mdot and little settling, is indicative of the turbulent enrichment of the inner disk atmosphere with large grains seen in other systems by \citet{sicilia-aguilar+07}.  It also suggests that there is a radial gradient in the maximum grain size of the upper layers, as the outer disk is best fit with 0.25$\mu$m grains.

\subsection{Ice-enhanced grain growth}
\label{iceab}
Although the silicate and graphite grains are small in the outer disk,  we can reproduce better the broader shape and flux of the SED from 50 to 70$\mu$m with a population of larger ice grains in the upper disk layers (Fig. \ref{varyice}b).  The 63$\mu$m peak is best fit by an $a_{max}$ of 15$\mu$m and a larger $\epsilon$, while a range of $a_{max}$ between 20 and 50$\mu$m can provide the bulk of the excess flux between 30 and 100$\mu$m.  The fact that we need ice grains two orders of magnitude larger than the silicate grains and with a different grain size distribution power law is consistent with simulations for ice enhancement of grain growth that take recondensation into account \citep[]{ks11}, if we assume that their ice-coated dust grain opacities can be approximated by our segregated ice grain opacities when $a_{max,i}$ is large.  The better fit to the 120$\mu$m region provided by the 50$\mu$m ice grains may indicate either a radial or vertical gradient in the ice grain size, which we do not yet implement in this version of our model. 

We can think of two potential explanations for why we might see an ice emission feature in this system and not (to the best of our knowledge) in other systems.  First, the disk of GQ Lup is highly unsettled ($\epsilon$=0.1-0.5).  This implies mixing of grain populations between the midplane and upper layers.  If icy grains are inclined to settle to the midplane or are formed there, we may see them because they are being turbulently lifted back to the disk surface.  Additionally, the lack of settling means that the location of the $A_V=3$ isocontour (which defines where ice ceases to be photodesorbed \citep{gh08}) is at a large enough vertical height in the disk that some ice should remain on the grains above the $\tau=1$ surface for 63$\mu$m emission.  Alternatively, if there are planetesimals in the disk, they would not appear to be large enough to clear a gap on the order of tens ofAU, as we do not see evidence for one in the SED \citep[although we note that this does not constrain smaller gaps;][]{espaillat+10}.  The presence of a massive, gaseous disk of $R_d$=30-100AU at the onset of the formation of Jovian planets has been simulated to scatter smaller planetesimals in the disk, populating high inclination orbits at $\sim$40AU \citep{kretke+12}.  If some of these icy planetesimals were to collide, it could generate a reservoir of icy grains in the upper layers \citep[e.g.][]{lisse+12}.  

\subsection{Evidence for localized heating in the outer disk?}

The presence of the 63$\mu$m feature is also of particular interest, as it appears only in crystalline ice, and the crystallization temperature of initially amorphous water ice is typically $>130$K \citep{baragiola03}.  Although the densities in our disk structure imply an ice sublimation temperature of at most 120 K  \citep{pollack+94}, it has been shown that ice deposited on grains at 160 K and subsequently cooled to 14 K maintains the 63$\mu$m opacity peak \citep{moore+01}.  This implies that GQ Lup has experienced significant heating and cooling events in the outer disk. 

Localized heating could also explain the annular crystalline distribution suggested by the last analysis in \S \ref{allanalysis}.  This distribution is not indicative of transport models, which we would expect to yield monotonically decreasing abundances, like those found by \citet{vanboekel+04} for some Herbig AeBe stars.  However, there are {\it in situ} mechanisms that could form crystals at tens ofAU.  For a typical T Tauri star, \citet{harkerdesch02} found that shocks from gravitational instabilities can anneal dust between 5 and 10AU, assuming an annealing temperature around 1200 K.  Crystalline dust could also be produced through collisions of parent bodies, as seen in older stars \citep[e.g.][]{lisse+12}.  Given the 3 Myr age of GQ Lup, it is not unfeasible that KBO-size planetesimals could have formed in the outer disk.

In conclusion, the mineralogical and disk structure properties of GQ Lup make it an interesting object with which to test theories of planetesimal formation. The grains have grown to micron size in the inner disk, while the outer disk is highly unsettled and shows signs of water ice in its PACS spectrum at sizes consistent with models of ice-enhanced grain growth.  The presence of crystalline ice and silicates at temperatures below their crystallization temperature and in annular distributions is indicative of a periodic {\it in situ} heating mechanism.  Follow-up studies should be conducted to confirm the disk size, resolve any radial structure at larger radii than those probed by our spectra, and probe the midplane grain size distributions.

\acknowledgments
M.K.M was supported by the National Science Foundation Graduate Student Research Fellowship under Grant No. DGE 0718128.  N.C and L.A. acknowledge support from NASA Origins grants NNX08AH94G.  C.E. was supported by the National Science Foundation under Award No. 0901947.


\begin{deluxetable}{cc}
\tabletypesize{\small}   
\tablewidth{0pt}
\tablecaption{Stellar and Model Properties}
\tablehead{\colhead{Parameter} & \colhead{Value}}
\startdata
 Stellar & (from SD08) \\
 \hline \\
$T_{eff}$ & 4060 K \\
$L_*$ & 1.0\Lsun  \\
$A_V$ & 0.5 mag \\
$M_*$ & 0.8\Msun  \\
$\dot{M}$ &  $5\times10^{-8}$\Msun$yr^{-1}$  \\
$i$ & 50\degr  \\
$d$ & 155$\pm$8pc  \\
 \hline \\
Wall &  \\
 \hline \\
  $T_{wall}$ & 1500 \\
 $R_{wall}$ & 0.19AU \\
$z_{wall}$ & 0.03AU \\
 $a_{max}$ & 3.0 \\
 composition & 100\% pyr, 40\% Fe\\
 silicates & 0.004 \\
 graphite & 0.0025\\
 \hline \\
 Disk &  \\
 \hline \\
$\alpha$ & 0.01$<\alpha\le$0.1 \\
 $\epsilon$ & 0.1$<\epsilon\le$0.5 \\
 $R_{d}$ & 50AU\\
 $a_{max}$ & 0.25$\mu$m (sil., graph.)\\
   & 15 - 70$\mu$m (H$_2$O ice) \\
 $M_{d}$ & 0.012\Msun \\
 silicate composition: & \\
 (radially constant) & 10\% forsterite, 10\% enstatite, \\
    &   80\% pyroxene \\
 (annular, suggestive) & 18\% forsterite, 15\% enstatite in outer 20-50AU \\
 abundances &  0.004 (sil) \\
    & 0.0025 (graf) \\
    & 2$\times$10$^{-3}$ (H$_2$O ice) 
\enddata
\label{sampletab}
\end{deluxetable}

\end{document}